\documentclass[a4paper, amsfonts, amssymb, amsmath, reprint, showkeys, nofootinbib, twoside, colorlinks=true, linkcolor=blue, citecolor=red, urlcolor=blue, aps, prd]{revtex4-2}
\usepackage[english]{babel}
\usepackage{esvect}
\usepackage{orcidlink}
\usepackage[utf8]{inputenc}
\usepackage[colorinlistoftodos, color=green!40, prependcaption]{todonotes}
\usepackage{ulem}
\usepackage{amsthm}
\usepackage{mathtools}
\usepackage{physics}
\usepackage{xcolor}
\usepackage{graphicx}
\usepackage[left=23mm,right=13mm,top=35mm,columnsep=15pt]{geometry} 
\usepackage{adjustbox}
\usepackage{placeins}
\usepackage[T1]{fontenc}
\usepackage{lipsum}
\usepackage{csquotes}
 
\usepackage{changes}

\usepackage[normalem]{ulem}

\def \vss{\vspace{14pt}}

\def\be{\begin{equation}}
\def\ee{\end{equation}}
\def\bea{\begin{eqnarray}}
\def\eea{\end{eqnarray}}

\def \D{{\mathcal{D}}}

\def \Mat {{\bf M}}
\def \H {\mathcal{H}}

\def \myS{{\mathcal{S}}}
\def \P{{\mathcal{P}}}
\def \a{\alpha}
\def \b{\beta}

\def \fmn{f_{\rm min}}
\def \fmx{f_{\rm max}}

\def \x{{\bf x}}
\def \y{{\bf y}}

\def \th {\tilde{h}}
\def \h{{\bf h}}
\def \Dh {\Delta \h}

\def \n{{\bf n}}

\def \e{{\bf e}}
\def \V{\mathcal{V}}
\def \hf{\frac{1}{2}}
\def \chisqrlns{\chi^2_{\rm{lens}}}

\begin{document}
\title{A $\chi^2$ statistic for the identification of strongly lensed gravitational waves from compact binary coalescences}

\author{{Sudhir Gholap$^{1}$}\orcidlink{0009-0009-4987-7114}}
    \email[sudhir.gholap@iucaa.in ]{}

\author{{Kanchan Soni$^{2,1}$}\orcidlink{0000-0001-8051-7883}}
    \email[ksoni01@syr.edu]{}
    
\author{{Shasvath J. Kapadia$^{1}$}\orcidlink{0000-0001-5318-1253}}
    \email[shasvath.kapadia@iucaa.in]{}

\author{Sanjeev Dhurandhar$^{1}$}
    \email[sanjeev@iucaa.in ]{}

\affiliation{$^1$Inter University Centre for Astronomy and Astrophysics, Post Bag 4, Ganeshkhind, Pune 411007}
\affiliation{$^2$Department of Physics, Syracuse University, Crouse Dr, Syracuse, NY 13210}

\date{\today} % Leave empty to omit a date

\begin{abstract}
Gravitational waves (GWs) emanated by stellar-mass compact binary coalescences (CBCs), and lensed by galaxy- or cluster-scale lenses, will produce two or more copies of the GW signal. In the geometric-optics limit, these signals will have similar phase evolution but differing amplitudes. Such lensing signatures are expected to be detected by the end of the LIGO-Virgo-KAGRA (LVK) collaboration's fifth observing run. In this work, we propose a novel $\chi_{\mathrm{lens}}^2$ statistic to segregate pairs of detected GW events as lensed or unlensed, using templates typically used in GW searches. The statistic is an application of the generalized $\chi^2$ discriminator, tailored to probe the similarity (or lack thereof) between the phase evolutions of two CBC signals. We assess the performance of $\chi_{\mathrm{lens}}^2$ on an in-part astrophysically motivated dataset of lensed and unlensed CBCs detectable in the fourth observing run, assuming a single LIGO-like detector at design sensitivity. Our results show that $\chi_{\mathrm{lens}}^2$ correctly identifies lensed events with efficiencies comparable to existing Bayesian and machine learning methods, while being orders of magnitude faster to evaluate than Bayesian methods. Moreover, the statistical properties of $\chi_{\mathrm{lens}}^2$ in stationary Gaussian noise are fully understood, in contrast to machine learning methods. $\chi_{\mathrm{lens}}^2$ can, therefore, be used to rapidly and accurately weed out the vast majority of unlensed candidate pairs and identify lensed pairs.
\end{abstract}

\keywords{Gravitational Waves, Gravitational Lensing, Compact Binaries}

\maketitle

\section{Introduction}
The LIGO-Virgo~\cite{TheLIGOScientific:2014jea, TheVirgo:2014hva} network of gravitational-wave (GW) detectors has observed $\sim 90$ compact binary coalescence (CBC) events in its first three observing runs (O1, O2, O3)~\cite{KAGRA:2021vkt}. Additional events were independently catalogued in~\cite{nitz_ogc3,venumadhav_2022_o3a,ogc4_Nitz_2023,mehta2024newbinaryblackhole,wadekar2023newblackholemergers} using publicly accessible LIGO-Virgo data~\cite{Vallisneri_2015,LIGOScientific:2019lzm}. The overwhelming majority of these are binary black holes (BBHs). Nevertheless, binary neutron star~\cite{TheLIGOScientific:2017qsa, Abbott:2020uma} and neutron star black hole mergers~\cite{LIGOScientific:2021qlt} have also been observed. The ongoing O4 is likely to more than double the number of detections, confirming that GW astronomy has well and truly arrived~\cite{Capote:2024rmo, abac2025gwtc}.

The detected CBCs have enabled some of the most unique tests of general relativity (GR) in the strong-field regime \cite{LIGOScientific:2021sio}. These include: a residuals test where the best fit GR waveform is subtracted out from the data, and the consistency of the residual with noise is assessed \cite{LIGOScientific:2016lio, LIGOScientific:2019fpa,LIGOScientific:2020ufj}; an inspiral-merger-ringdown consistency test, where source parameters extracted from low and high-frequency parts of the signal, assuming a GR template waveform, are compared \cite{Ghosh:2016qgn,
Ghosh:2017gfp}; a test that looks for deviations in Post Newtonian coefficients of the GW signal as a means to investigate the consistency of the inspiral with GR \cite{Blanchet:1994ez, blanchet1994signal, Arun:2006hn, Arun:2006yw,
Yunes:2009ke, Mishra:2010tp, Li:2011cg, Li:2011vx}; and propagation tests that compare the speed of GWs with the speed of light \cite{LIGOScientific:2018dkp}, and look for signatures of velocity dispersion due to a non-zero graviton mass \cite{Will:1997bb}.

Gravitational lensing of GWs promises to provide yet another unique test of GR. This phenomenon occurs when GWs encounter massive structures. Strong lensing of GWs happens when the wavelength of the passing GW signal is much smaller than the Schwarzschild length scale of the lens mass, i.e, $\lambda_{\mathrm{GW}} \ll 2\mathrm{G}M_{\mathrm{lens}}/\mathrm{c}^2$. The effect of lensing on GWs can be understood by calculating Kirchhoff's diffraction integral. In particular, when GWs from stellar mass CBCs encounter galaxies or clusters, strong lensing in the geometric optics regime will ensue. Under the geometric optics approximation, only the stationary points of the time-delay surface contribute to the diffraction integral. Thus, the integral reduces to a sum over contributions from the stationary points, given by:
\begin{equation}
    F(f) = \sum_{j}^{} |\mu_j|^{1/2} \mathrm{exp}\left[ i 2\pi f t_j - i \pi \, \mathrm{sign}(f) \, n_j \right]\,.
\end{equation}

Here, $\mu_j, t_j ~\mathrm{and}~ n_j$ denote the magnification, time of arrival, and Morse index associated with the $j^{\text{th}}$ image, respectively. The observed signal is the product of $F(f)$ and the unlensed signal, resulting in multiple temporally resolved copies of the GW signal being detected by the detector network. The image time delays span minutes to months for galaxy-scale lenses \cite{Ng2018,
li2018gravitational, oguri2018effect}, and weeks to years for cluster-scale
lenses \cite{smith2018if, smith2020massively,Smith:2019qsv, robertson2020does, ryczanowski2020building}. These copies will have identical phase evolutions, but differing amplitudes (due to (de)magnification) as well as a constant (Morse) phase difference of $0, \pi/2$, or $\pi$ for Type I ($n_j=1$), Type II ($n_j=1/2$), and Type III ($n_j=1$) \cite{wang1996, Dai:2017huk, ezquiaga2021}. On the other hand, if these GWs encounter lenses, such as intermediate-mass black holes, whose Schwarzchild radii are comparable to the wavelength of the GWs, a single modulated GW image will be produced, exhibiting interference, diffraction, and beating patterns \cite{deguchi1986,nakamura1998, takahashi2003wave, cao2014, lai2018,
christian2018, dai2018, diego2020, Chan:2024qmb}. Gravitational lensing will enable additional powerful tests of GR \cite{baker2017, collett2017, fan2017, goyal2021,
ezquiaga2020}; provide precise measurements of cosmological parameters \cite{sereno2011cosmography,liao2017precision, cao2019direct,
li2019constraining, hannuksela2020localizing, jana2023, Jana:2024dhc}; drastically enhance GW early warning \cite{Magare:2023hgs}; enable constraints on the fraction of dark matter as massive compact halo objects \cite{Basak:2021ten, Barsode:2024wda}, as well as on the mass of the warm dark matter particle \cite{Jana:2024uta}; probe properties of the lens such as electric charge \cite{Deka:2024ecp}, and probe the proper motion of isolated Galactic neutron stars \cite{Basak:2022fig}. 

The LVK Collaboration has conducted multiple searches for signatures of GW lensing during its first three observing runs. However, no confirmed detection of gravitational lensing of GWs has been reported \cite{Hannuksela:2019kle, LIGOScientific:2021izm, LIGOScientific:2023bwz}. Based on lensing rate estimates, it is expected that the first detection of strongly lensed GWs will occur before the end of O5 \cite{wierda2021beyond, Xu:2021bfn}. In anticipation of this exciting possibility, a number of search methods have been proposed. These include: 
\begin{itemize}
    \item A posterior overlap statistic that measures the degree of overlap between posteriors of the source parameters, inferred from the data surrounding each of the events in a candidate pair \cite{Haris:2018vmn}. An upgrade to this statistic, by self-consistently incorporating prior astrophysical information, was recently provided \cite{Barsode:2024zwv}.
    \item A machine learning (ML) based method that uses time-frequency maps and localization skymaps of the candidate pair to provide a probability of class membership (lensed or unlensed) \cite{Goyal:2021hxv, Magare:2024wje}.
    \item An approximate method that combines localization skymaps with the Bhattacharya distance between detector-frame mass posteriors, approximated as Gaussians, of the two events in a candidate pair \cite{Goyal:2023lqf}.
    \item A method that compares the GW phase of the two events in the pair at a fixed reference frequency \cite{Ezquiaga:2023xfe}, and another that cross-correlates data surrounding each of the events in the pair \cite{Chakraborty:2024net}.
    \item Comprehensive Bayesian methods that construct a joint likelihood from the data surrounding each of the events in the candidate pair \cite{liu2021identifying, Janquart:2022wxc, Janquart:2023osz, Lo2023, 2024arXiv241201278B}. 
\end{itemize}

Each of these methods has its unique strengths and shortcomings. For example, the posterior overlap statistic, as well as comprehensive Bayesian methods require large-scale parameter estimation runs that could take anywhere from hours to weeks, per event candidate pair\footnote{It may be argued that the posterior overlap statistic uses existing posteriors on the individual detected GW events and does not require a separate large-scale parameter estimation run. Nevertheless, not all GW events have PE posteriors readily available, such as subthreshold events.}. On the other hand, rapid methods such as those based on ML require carefully constructed training sets and suffer from issues pertaining to robustness and interpretation. The approximate method mentioned above, while rapid, suffers from the fact that the Gaussian approximation to the posterior may not be valid in real noise for low-significance events. 

In this work, we propose a template-based method to rapidly identify a pair of events as lensed or unlensed. The method applies the generalized $\chi^2$ statistic proposed in \cite{dhurandhar2017} and tunes it to assess the phase-evolution-consistency between quasicircular and quadrupolar GW signals in a candidate lensed pair\footnote{The generalized $\chi^2$ was also used in the context of segregating signals from glitches in Ref. \cite{Joshi:2020eds}.}. In particular, using the trigger template\footnote{A trigger template refers to a coincident trigger whose ranking statistic exceeds a predefined threshold.}, as well as neighboring ones associated with the louder of the two events in a candidate pair, a vector space orthogonal to this neighborhood is constructed. The strain data pertaining to the second (weaker) signal is then projected onto the orthogonal vector space. 

In the absence of noise, a lensed pair would yield a zero projection, while an unlensed pair would yield a projection proportional to the second event's signal-to-noise ratio (SNR) squared. In the presence of stationary Gaussian noise, the projection for the lensed pair will be a $\chi^2$ random variable, whose mean will depend on the number of degrees of freedom of $\chi^2$. On the other hand, for an unlensed pair, the mean will depend on the SNR squared, in addition to the degrees of freedom. 

We assess the performance of $\chi^2_{\mathrm{lens}}$ on an in-part astrophysically motivated dataset, consisting of lensed and unlensed pairs in an approximate ratio of $1:1600$. These signals are injected into Gaussian noise, assuming an O4-like (design sensitivity) power spectral density (PSD) and a single detector configuration, with the constraint that the weaker signal in the candidate pair has an optimal SNR $\geq 8$. We find that our method successfully distinguishes lensed and unlensed pairs with True Positive Probability (TPP) at a low False Positive Probability (FPP), comparable to the posterior overlap statistic. Moreover, we study the performance of $\chi^2_{\mathrm{lens}}$ as a function of the SNR of the candidate pairs, as well as their duration within the sensitivity band of the detector. As expected, we find that increasing the SNR and in-band duration improves the performance of $\chi^2_{\mathrm{lens}}$. Overall, our results demonstrate that $\chi^2_{\mathrm{lens}}$ statistic provides a rapid, accurate, and readily interpretable method to identify strongly lensed gravitational wave pairs. 

The rest of the paper is organized as follows. Section~\ref{Sec:Methods} provides a summary of the posterior overlap statistic and briefly introduces the Unified $\chi^2$ formalism. Section \ref{sec:chisquare_lens} describes the construction of the $\chisqrlns$ statistic developed in this work and examines its properties. Section \ref{Sec:Results} outlines the results demonstrating the performance of $\chi_{\mathrm{lens}}^2$. Finally, Section~\ref{Sec:Conclusions} summarizes the paper and discusses the scope for future work.
We also discuss the unified $\chi^2$ formalism in more detail in Appendix \ref{append_unified_chisq}.

\section{Methods}\label{Sec:Methods}
In this section, we summarize the posterior overlap statistic, which serves as a benchmark for comparison. We also briefly introduce the unified $\chi^2$ of \cite{dhurandhar2017}, and provide details in Appendix~\ref{append_unified_chisq}.

\subsection{The Posterior Overlap Statistic}

Consider a detector strain time series $x(t)$, containing a GW signal $A\, \hat{h}^{\rm{s}}(t, \vec{\theta})$ with amplitude (optimal SNR) $A$, source parameters $\vec{\theta}$, and noise $n(t)$. The noise $n(t)$ is usually modeled as a realization of stationary Gaussian noise with zero mean and a power spectral density $S_{\rm{n}}(f)$. The strain is then a simple superposition of the two,
\begin{equation}\label{eq:data}
    x(t) = n(t) + A\,\hat{h}^{\rm{s}}(t, \vec{\theta}) \,.    
\end{equation}
The presence of $A\, \hat{h}^{\rm{s}}(t, \vec{\theta})$ in $x(t)$ is probed by evaluating the matched-filter SNR $\rho = (x|\hat{h}^{\rm{t}})$. Here, $\hat{h}^{\rm{t}}$ is a normalized GW template, and the inner product $(\cdot | \cdot)$ is defined as:
\begin{equation}
    (a | b) = 4 \Re \int_{\fmn}^{\fmx} \frac{\tilde{a}^*(f)\tilde{b}(f)}{S_{\rm n}(f)}df \,,
\label{eq:scalar_gw}    
\end{equation}
where $a(t)$ and $b(t)$ are two time series, and $\tilde{a}$ and $\tilde{b}$ are their Fourier transforms. If $\hat{h}^{\rm{t}}$ is an exact representation of the signal in the data $\hat{h}^{\rm{s}}(t, \vec{\theta})$, then $\rho$, on average, will equal the so-called optimal SNR $\langle \rho \rangle = \rho_{\mathrm{opt}} = (A\,\hat{h}^{\rm{s}} | \hat{h}^{\rm{t}}) = A$.

Inferring the parameters $\vec{\theta}$ requires extensive parameter estimation runs. During these runs, the GW posterior distribution is sampled:
\begin{equation}
    p(\vec{\theta}|x) = \frac{p(\vec{\theta})p(x | \vec{\theta})}{p(x)}\,,
\end{equation}
where $p(x | \vec{\theta}) \propto \exp[-(x - A\,\hat{h}^{\rm{t}} | x - A\,\hat{h}^{\rm{t}})/2]$ is the likelihood, $p(\vec{\theta})$ is the prior, and $p(x)$ is a normalization constant called the evidence. 

The posterior overlap statistic acts as a discriminator between lensed and unlensed candidate pairs of detected GW events. It exploits the fact that lensed events share identical phase evolutions, and therefore their intrinsic parameters in the detector frame must also be identical. Consequently, the degree of overlap between the posteriors of the intrinsic parameters of the two events could be indicative of the lensed nature of the pair. 

Formally, the posterior overlap statistic is defined as follows. Consider two data streams $x_1(t)$ and $x_2(t)$, known to contain GW signals $A_1\,\hat{h}^{\rm{s_1}}(t)$ and $A_2\,\hat{h}^{\rm{s_2}}(t)$, respectively. Let $\mathcal{H}_{\rm L}$ be the hypothesis that $A_{1,2}\,\hat{h}^{\rm{s_{1,2}}}(t)$ are lensed copies, and let $\mathcal{H}_{\rm U}$ be the hypothesis that the two signals are unrelated (i.e., unlensed). Assuming no prior information on which hypothesis is preferred, a Bayesian discriminator can be constructed from the ratio of the evidence of the joint dataset $\lbrace{x_1(t), x_2(t) \rbrace}$ conditioned on each of the two hypotheses \cite{Haris:2018vmn}
\begin{equation}\label{eq:blu}
    \mathcal{B}^{\rm L}_{\rm U} = \frac{p(x_1, x_2|\mathcal{H}_{\rm L})}{p(x_1, x_2|\mathcal{H}_{\rm U})} = \int \frac{p(\vec{\theta}|x_1)p(\vec{\theta}|x_2)}{p(\vec{\theta})}d\vec{\theta} \,.
\end{equation}
The posterior overlap statistic \cite{Haris:2018vmn} is then simply a Bayes factor, which can be constructed using the individual parameter posteriors of the two GW events and the prior used in the construction of the posteriors.

\subsection{A brief introduction to the unified $\chi^2$ formalism}
% \sjk{Notational Inconsistency: $i$ and $\iota$ are being used interchangeably (see Eq. 1). Needs to be fixed.}
The general framework for $\chi^2$ discriminators has been described in~\cite{dhurandhar2017}, where various $\chi^2$ discriminators are unified into a single discriminator, which can be appropriately termed as the
{\it unified} $\chi^2$.
In this framework, a data stream $x(t)$ defined over a time interval $[0,\, T]$ is viewed as a vector $\x$ denoted in boldface. Such data streams form a vector space $\D$ of a large number of dimensions equal to the number of sample points in the data stream. The scalar product, defined in Eq.~\eqref{eq:scalar_gw}, turns this space into a Hilbert space. The detector noise $n(t)$ is a stochastic process defined over the time segment $[0,\,T]$ and is now written as a random vector $\mathbf{n}$. %Eq. (\ref{eq:data}) can be written as $\mathbf{x} = \mathbf{n} + A \widehat{\mathbf{h}^{\mathrm{s}}}$, where $\widehat{\mathbf{h}^{\mathrm{s}}}$ is a normalised signal vector and $A$ the amplitude. 
Further details of the unified $\chi^2$ formalism can be found in Appendix \ref{append_unified_chisq}.

A waveform could have an arbitrary phase in the time domain, which can be written as a linear combination of two orthogonal templates $\h_0$ and $\h_{\pi/2}$. In the Fourier domain, these two waveforms are simply related by a factor of the complex number $i$. Taking only the positive frequency components contains full information of the real data, and this is explicitly shown in Appendix \ref{append_A}. The complex approach is useful because the waveforms span a 2-dimensional space parametrized by the coalescence phase. We, therefore, consider only the positive frequency part $f > 0$ of the signal as well as of the data. This gives us complex vectors that again belong to a Hilbert space $\H$, which is essentially $L^2 [\fmn, \fmx]$ with a weighted measure defined by $d \mu = df/S_{\rm n}(f)$. On $\H$, we define the scalar product of two vectors $\mathbf{g}_1, \mathbf{g}_2 \in \mathcal{H}$ as,
\be
(\mathbf{g}_1, \mathbf{g}_2) = 4 \int_{\fmn}^{\fmx} \frac{\tilde{g_1}^* (f) ~ \tilde{g_2} (f)}{S_{\rm n} (f)} df \,.
\label{eq:scalar_u}
\ee

We have scaled the usual scalar product by a factor of 4 so that the norms of any vector, as computed by either scalar product, agree. This is shown in Eq.~\eqref{eq:norm} in Appendix \ref{append_A}. The scalar product in Eq.~\eqref{eq:scalar_u} is related in a simple way to the one defined in Eq.~\eqref{eq:scalar_gw}. In Appendix \ref{append_A}, this relation is explicitly deduced. The scalar product is complex in general, satisfying $(\mathbf{g}_2, \mathbf{g}_1) = (\mathbf{g}_1, \mathbf{g}_2)^*$, which means the order of the vectors appearing in the scalar product is important and must be taken into account while performing computations. We assert that $\H$ is isomorphic to $\D$. The entire procedure for constructing the $\chi^2$ can be followed through as in the case of $\D$, where now we deal with complex vectors and use the scalar product as defined by Eq.~\eqref{eq:scalar_u}. The resulting $\chi^2$ is, of course, real and positive, although it has an even number of (real) degrees of freedom.

\section{$\chi_{\mathrm{lens}}^2$: Construction, Properties and Interpretation}
\label{sec:chisquare_lens}
\subsection{The recipe}
\label{subsec:recipe}

In the context of strong-lensing, consider two strain data segments $\mathbf{x}_1$ and $\mathbf{x}_2$, containing signals $\widehat{\mathbf{h}}^{\rm s_1}$
and $\widehat{\mathbf{h}}^{\rm s_2}$, with amplitudes $A_1$ and $A_2$ $(A_1 > A_2)$ respectively and buried in Gaussian noise. Since the signals arrive at different times, the realisations of this noise will in general differ and we label them as $\mathbf{n}_1 ~\mathrm{and}~ \mathbf{n}_2$. Note that their statistical properties are the same. Then we have 
\begin{subequations}
   \begin{align}
    \mathbf{x}_1 &= A_1 \cdot \widehat{\mathbf{h}}^{\rm s_1} + \mathbf{n}_1\,,  \\
    \mathbf{x}_2 &= A_2 \cdot \widehat{\mathbf{h}}^{\rm s_2} + \mathbf{n}_2\,.
    \end{align} 
\end{subequations}
Here, the symbol $~\widehat{}~$ over a vector implies that the vector under consideration has a unit norm with respect to the inner product defined in Eq.~\eqref{eq:scalar_u}. The signals $\widehat{\mathbf{h}}^{\rm s_1}$ and $\widehat{\mathbf{h}}^{\rm s_2}$, when strongly lensed, will have the same intrinsic parameters $\vv{\theta}$  along with a relative difference in their Morse phases. Note that, throughout this work, we consider non-spinning CBC sources that emit dominant-mode signals as proof-of-concept. However, the method we propose is expected to work in general as long as the images exhibit phase-evolution consistency and should also be applicable to spinning templates.

Matched filtering of the strain data $\mathbf{x}_1$ and $\mathbf{x}_2$ identifies templates $\widehat{\mathbf{h}}(\vv{\theta_1})=\widehat{\mathbf{h}}^{\rm t_1}$ and $\widehat{\mathbf{h}}(\vv{\theta_2})=\widehat{\mathbf{h}}^{\rm t_2}$, which maximize the SNR, from the template bank $\mathcal{B}$, respectively. We aim to construct a space $\mathcal{S}$ orthogonal to the space containing the trigger $\widehat{\mathbf{h}}^{\rm t_1}$ corresponding to the higher-SNR signal in the pair under consideration and the templates lying in its neighborhood, which is our main goal for this section. This is done to get a space approximately orthogonal to the unknown signal $\widehat{\mathbf{h}}^{\rm s_1}$.
A desirable property of this orthogonal space $\mathcal{S}$ is that the signal $\widehat{\mathbf{h}}^{\rm s_2}$(which will have the same phase evolution as $\widehat{\mathbf{h}}^{\rm s_1}$ when strongly lensed), should have a small projection onto the orthogonal space. Also, the signal $\widehat{\mathbf{h}}^{\rm s_2}$ is expected to have a large projection onto the orthogonal space when the pair under consideration is unlensed. Now to construct $\mathcal{S}$, we start with the trigger template $\widehat{\mathbf{h}}^{\rm t_1}$ and 
 define the set of neighborhood templates $\left\{ \widehat{\mathbf{h}}_{\alpha}\right\}$ selected from the template bank $\mathcal{B}$ by setting a match $\mathbb{M}$ (complex inner product maximized over time $t_{\rm c}$ and phase $\phi_{\rm c}$ at coalescence) threshold $\mu$ (e.g., 0.97) between trigger template $\widehat{\mathbf{h}}^{\rm t_1}$ and template $\widehat{\mathbf{h}}^{\rm t}$  from the bank as follows:
\begin{equation}
    \left\{ \widehat{\mathbf{h}}_{\alpha}\right\} = \left\{  \widehat{\mathbf{h}}^{\rm t} \in \mathcal{B} | \left( \widehat{\mathbf{h}}^{\rm t}\,, \widehat{\mathbf{h}}^{\rm t_1}\right)_{{\rm max}(t_{\rm c}, \phi_{\rm c})} \geq \mu  \right\}\,.
\end{equation}
The index $\a$ runs from $1$ to $J$, where $J$ is the number of templates in $\left\{ \widehat{\mathbf{h}}_{\alpha}\right\}$. Clearly, the trigger template $\widehat{\mathbf{h}}^{\rm t_1}$ is included in this collection. Now, certain regions of the parameter space in the template bank might be overdense\footnote{This is particularly the case for the high mass region of the parameter space if one is using a geometric bank.} in the number of templates. If the trigger template $\widehat{\mathbf{h}}^{\rm t_1}$ 
 happens to lie in such a region of parameter space, we might encounter a situation where we have a large number of templates, say $\sim 100$, in the neighborhood of the trigger template.  This could make our computation of the $\chi^2$ cumbersome.
\par

In order to deal with this kind of situation, we use Singular Value Decomposition (SVD). We construct a matrix $\Mat$ as follows. We consider the templates $\widehat{\mathbf{h}}_{\alpha}$ in the frequency domain and consider only positive frequency components, with frequencies between $\fmn$ and $\fmx$. Then $\widehat{\mathbf{h}}_{\alpha}$ can be thought of as complex vectors in the interval $[\fmn, \fmx]$. These vectors are in $\H$. Since we are in a complex space, we must use the scalar product defined by Eq.~\eqref{eq:scalar_u} on $\H$ rather than the one given by Eq.~\eqref{eq:scalar_gw}. 
We need to whiten the row vectors  $\widehat{\mathbf{h}}_{\alpha}$ before applying the SVD. This is because the standard SVD algorithm uses the Euclidean scalar product without the PSD. This results in the matrix $\Mat$ consisting of $J$ row vectors, which span a $ J$-dimensional complex vector space $\mathcal{W}_{J}$, assuming the templates to be linearly independent. But since $J$ may be large, we use the SVD to whittle down the row dimension of $\Mat$ and get the best low-dimensional approximation to $\mathcal{W}_{J}$. This approximate space we call $\mathcal{W}$. The SVD implementation is described in detail in Section~\ref{Subsubsec:SVD}. Further, the SVD also gives us an orthonormal set of vectors $\lbrace \widehat{\mathbf{w}}_{\alpha} \rbrace$ - the right singular vectors - which then form an orthonormal basis of $\mathcal{W}_{J}$. When $J$ is large, it is usually possible to make the dimension $j$ of $\mathcal{W}$ much less than $J$. We then unwhiten the singular vectors corresponding to the most significant singular values by multiplying them with $\sqrt{S_{\rm n}(f)}$ to obtain a set $\lbrace \widehat{\mathbf{v}}_{\alpha} \rbrace $ which spans a space $\V$.

The next step is to obtain a space $\mathcal{S}$ orthogonal to $\V$, which will then be at least approximately orthogonal to the linear span of the neighborhood templates, namely, the collection $\widehat{\mathbf{h}}_{\alpha}$. Given a template with coalescence phase equal to zero, say $\widehat{\mathbf{h}}_{0}$, we can obtain the template with phase $\pi/2$ by just multiplying it by $i$, since we are only considering positive frequencies, that is, $\tilde{\rm h} _{\pi/2} (f) = i \tilde{\rm h} _{0}(f)$ or in vector notation, $\widehat{\mathbf{h}}_{\pi/2} = i \widehat{\mathbf{h}}_{0}$. Both $\widehat{\mathbf{h}}_{0}$ and $\widehat{\mathbf{h}}_{\pi/2}$ are regarded as complex vectors. In the complex space, the two vectors are complex multiples of each other. So, we need to consider only one of them, say $\widehat{\mathbf{h}}_{0}$. We, therefore, proceed to construct a complex vector orthogonal to the space $\V$. We do this by subtracting out the component of $\widehat{\mathbf{h}}_{0}^{\rm t_2}$ which projects onto the space $\V$. Therefore, we have,

\begin{subequations}
   \begin{align}
    \Dh &= \widehat{\mathbf{h}}_{0}^{\rm t_2} - \mathbf{v}\label{eq:deltah_def} \\
    &= \widehat{\mathbf{h}}_{0}^{\rm t_2} - \sum_{\alpha = 1}^{j} \left ( \widehat{\mathbf{v}}_{\alpha}, \widehat{\mathbf{h}}_{0}^{\rm t_2} \right ) \widehat{\mathbf{v}}_{\alpha}\,,
\end{align} 
\end{subequations}

where $\mathbf{v} = \sum_{\alpha = 1}^{j} \left ( \widehat{\mathbf{v}}_{\alpha}, \widehat{\mathbf{h}}_{0}^{\rm t_2} \right ) \widehat{\mathbf{v}}_{\alpha}$ is the component of $\widehat{\mathbf{h}}_{0}^{\rm t_2}$ parallel to $\mathcal{V}$.
Thus, by construction, $\Dh$ is orthogonal to $\V$, and it spans the space  $\myS$ of one complex dimension. This corresponds to the two real dimensions of the two phases $0, \pi/2$. We note that $\Dh$ is a complex vector accommodating both phases. However, it is not normalized. The normalized vector is given by
\be
\widehat{\Delta \mathbf{h}} = \frac{\Dh}{\| \Dh \|}  \,,
\ee
where the norm is computed from the scalar product in Eq.~\eqref{eq:scalar_u}. The corresponding complex correlation is the projection of the data vector $\mathbf{x}_2$ onto $\widehat{\Delta \mathbf{h}}$, that is, \break $\Delta C = (\mathbf{x}_2, \widehat{\Delta \mathbf{h}})$. The $\chi^2$ statistic is given by
\be
\chi^2_\mathrm{lens} = | \Delta C |^2  \,.
\ee
In Gaussian noise, $\chi^2_\mathrm{lens}$ has a non-central $\chi^2$ distribution with two degrees of freedom in general (see Section \ref{Subsubsec:effect_of_mismatch}).
\vss

\subsection{Singular Value Decomposition}
\label{Subsubsec:SVD}

In this section, we describe the SVD and its necessity in our work. As discussed in the previous section, certain trigger templates may have a large number of templates in their corresponding neighborhood, many of which are redundant. The SVD algorithm is a powerful tool for reducing the number of vectors needed while preserving the essential characteristics of the space spanned by the neighborhood template vectors $\widehat{\mathbf{h}}_{\a}$.

To achieve this reduction, we proceed with the following steps:

\begin{itemize}
    \item We first prepare the SVD input matrix $\Mat$ by taking frequency-domain normalized templates $\widehat{\mathbf{h}}_{\a}$, which are defined in the positive frequency range $\left[ f_{\rm min}, f_{\rm max} \right]$. Since we are using the scalar product of two vectors defined in Eq.~\eqref{eq:scalar_u}, which is an inner product weighted inversely by $S_{\rm n}(f)$, we need to whiten the templates by dividing each of them by $\sqrt{S_{\rm n}(f)}$. This is required because the SVD algorithm uses the Euclidean inner product\footnote{The Euclidean inner product of two vectors $\mathbf{p\mathrm{,}q} \in \mathbb{R}^n$ is defined as $\left( \mathbf{p\mathrm{,}q} \right) = p_1q_1 + p_2q_2 + ...+p_nq_n$.}.

    \item The SVD input matrix $\Mat$ is now defined as the matrix whose $\a$th row is the complex vector $\widehat{\mathbf{h}}_{\a}(f)/\sqrt{S_{\rm n}(f)}$. Thus, the resulting matrix will be ($J\times K$) dimensional, where $J$ and $K$ are the number of templates and the number of frequency sample points of a vector, respectively.

    \item The SVD algorithm factorizes the matrix $\mathbf{M}$ as a product of 3 matrices
    \begin{equation}
        \Mat_{(J\times K)} = \mathbf{U}_{(J\times R)}\mathbf{\Sigma}_{(R\times R)} \mathbf{V}^{\dag}_{(R\times K)}\,.
    \end{equation}
    Here $R \leq \min (J, K)$ is the `rank' or the number of linearly independent columns/rows of the input matrix $\Mat$. $\mathbf{\Sigma}$ is a diagonal matrix that contains the real and positive singular values ($\sigma$) sorted in descending order. Also, $\mathbf{U}$ and $\mathbf{V}$  are complex unitary matrices whose columns consist of the left and right singular vectors, respectively, forming an orthonormal basis for the column and row spaces of $\Mat$.

    \item Although the row vectors of $\mathbf{V}^{\dag}$ form an orthonormal basis for the row space of $\Mat$, many of them may be redundant on account of the overdensity of templates in certain regions of the parameter space. We aim to retain only the most significant basis vectors for our analysis while preserving the essential properties of the row space of $\Mat$. Specifically, we seek the best low-dimensional approximation to the row space of $\Mat$. This is achieved by selecting the first $j$ row vectors of $\mathbf{V}^{\dag}$ corresponding to the largest singular values. Given that the Frobenius norm of the SVD input matrix is $\lvert| \Mat \rvert|_{\rm Frob} = \sum_{\alpha=1}^{J} \sigma_{\alpha}^{2}$,
    \begin{equation}
        \sum_{\alpha=1}^{j} \sigma_{\alpha}^{2} \geq \zeta \lvert| \Mat \rvert|_{\rm Frob}, \quad  j \leq J\,.
    \end{equation}
    To ensure that the essential characteristics of the vector space are preserved, we set $\zeta = 99.9\%$ in our analysis.
    
    \item Finally, we unwhiten each significant basis vector by multiplying each frequency sample point of a vector by $\sqrt{S_{\rm n}(f)}$. The set of vectors, $\{\widehat{\mathbf{v}}_{\alpha}\}$, forms an orthonormal basis with respect to the inner product defined in Eq.~\eqref{eq:scalar_u}. This basis's linear span is $\mathcal{V}$, which best approximates the space spanned by $\left\{ \widehat{\mathbf{h}}_{\alpha}\right\}$.
    
\end{itemize}

\subsection{$\chi^2_{\mathrm{lens}}$ for lensed and unlensed pairs}

\begin{figure*}[h!t]
    \centering
    \includegraphics[width=\textwidth]{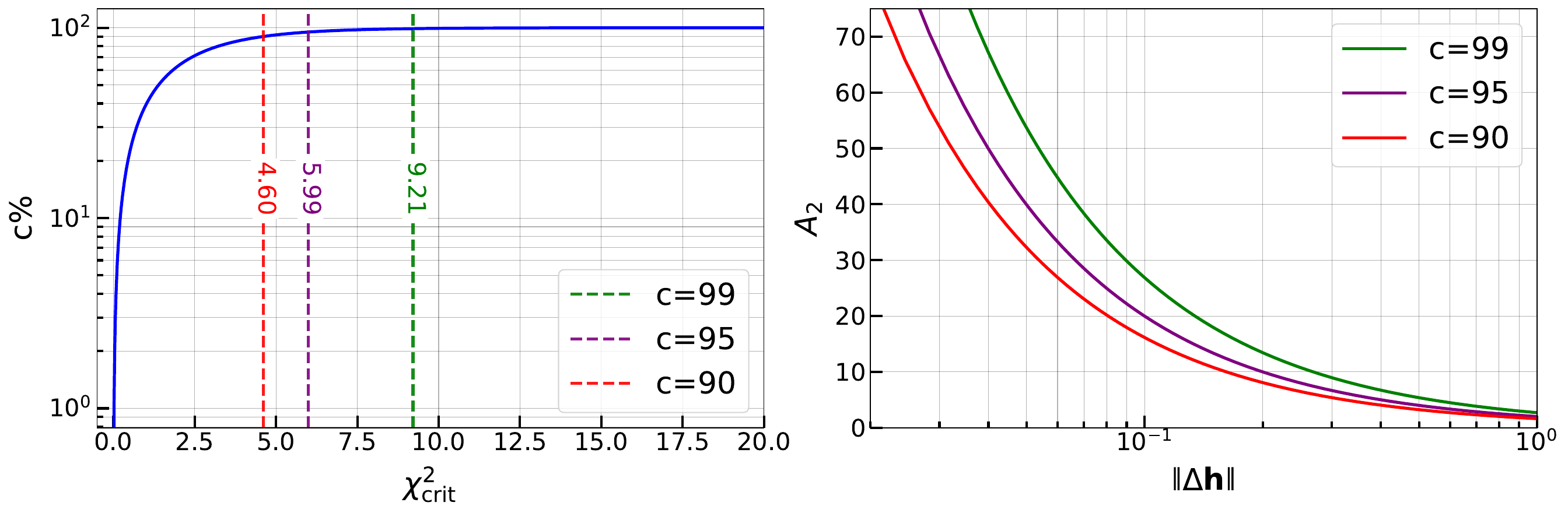}
    \caption{(Left) Confidence $c\%$ vs. $\chi^{2}_{\rm crit}$. Any pair of events having $\chi^{2}_{\rm lens} \leq \chi^{2}_{\rm crit}|_{c\%}$ could be classified as unlensed with $c\%$ confidence, assuming $\chi^{2}_{\rm lens}$ for lensed cases is a central $\chi^2$ ($\lambda = 0$) distribution. (Right) The minimum amplitude ($A_2$) required to classify a truly unlensed pair as unlensed vs. $ \| \Delta \mathbf{h}\|$ for varying  $c\%$. Smaller projection of the template $\hat{\mathbf{h}}_{0}^{\rm t_2}$ onto the orthogonal space results in $ \| \Delta \mathbf{h}\|$ approaching zero, necessitating larger amplitudes for unlensed classification at the same confidence level.}
\label{fig:chisq_cut}    
\end{figure*}

In this section, we investigate the distribution of $\chi^2_{\mathrm{lens}}$. We find that it possesses a non-central $\chi^2$ distribution in general. We evaluate its mean value and variance for lensed and unlensed pairs. For now, we assume that the template associated with the second signal is a perfect representation of that signal. In the following section, we investigate the effect of a small mismatch between signal and template.

To start with, it is straightforward to see that
\begin{equation}
    \left(\widehat{\mathbf{h}}_{0}^{\rm t_2}, \Delta \mathbf{h}\right) = 1 - \sum_{\alpha = 1}^{j}|\mathcal{M}_{\alpha}|^2\,,
\label{eq:proj}    
\end{equation}
where
\begin{equation}
\mathcal{M}_{\alpha} \equiv \left ( \widehat{\mathbf{h}}_{0}^{\rm t_2} \,,\widehat{\mathbf{v}}_{\alpha} \right )\,.
\end{equation}
Eq.~\eqref{eq:proj} also gives the square of the norm of $\Dh$ because from Eq.~\eqref{eq:deltah_def}, $(\mathbf{v}, \Dh) = 0$ and the result follows. Thus,

\begin{subequations}
   \begin{align}
    \|\Delta \mathbf{h}\| &= \sqrt{\left ( \Delta \mathbf{h}, \Delta \mathbf{h} \right)} = \sqrt{1 -\sum_{\alpha = 1}^{j}|\mathcal{M}_{\alpha}|^2},  \\
    \|\mathbf{v}\| &= \sqrt{\left( \mathbf{v}, \mathbf{v} \right)} =\sqrt{\sum_{\alpha = 1}^{j}|\mathcal{M}_{\alpha}|^2}. 
    \end{align} 
\end{subequations}
Assuming a perfect match between signal and template, a signal with an arbitrary phase is $\widehat{\mathbf{h}}^{\rm s_2} = \widehat{\mathbf{h}}_{0}^{\rm t_2}e^{i\phi}$. The absolute value of the complex correlation, in the absence of noise, is then

\begin{align}
    |\Delta C| = \left|\left (A_2 \widehat{\mathbf{h}}^{\rm s_2}, \widehat{\Delta \mathbf{h}} \right) \right | &= \left| e^{-i\phi} A_2\left( \widehat{\mathbf{h}}^{\rm t_2}_{0}, \widehat{\Delta \mathbf{h}} \right) \right | \nonumber \\
    &= A_2  \|\Delta \mathbf{h}\|\,.
\end{align}

In the presence of stationary Gaussian noise, $\Delta C$ is a Gaussian random variable with mean $\left( A_2 \widehat{\mathbf{h}}^{\rm s_2}, \widehat{\Delta \mathbf{h}} \right)$ and unit variance for both degrees of freedom. The correlation modulus squared then becomes a non-central $\chi^2$ random variable, $\chi^2_{\rm{lens}}$ having two degrees of freedom\footnote{$\chi^2_{\rm{lens}}$ has two degrees of freedom because $\Delta C$ is the sum of two unit-variance Gaussian random variables, one each for the real and imaginary part.} with noncentrality parameter $\lambda = \left| \left( A_2 \widehat{\mathbf{h}}^{\rm s_2}, \widehat{\Delta \mathbf{h}} \right) \right|^2$. Its mean value is

\begin{align}
    \langle \chi^2_{\rm{lens}} \rangle &= \lambda + 2 \nonumber\\
    &= |A_2|^2 \|\Delta \mathbf{h}\|^2 + 2\,.
\end{align}

The variance $\sigma_{\chisqrlns}^{2}$ (not to be confused with the singular values described in Section \ref{Subsubsec:SVD}) of $\chi^2_{\rm{lens}}$ is then\footnote{Variance of a non-central $\chi^2$ distribution with $p$ degrees of freedom having non-centrality parameter $\lambda$ is $2(p + 2\lambda)$.},

\begin{equation}
    \sigma_{\chisqrlns}^2 = 4(1 + \lambda)\,.
\end{equation}

When a pair of events is lensed, the second signal can be expressed, to a very good approximation, as a linear combination of $\widehat{\mathbf{v}}_{\alpha}$. Indeed, if the second template perfectly matches one of the templates neighboring the first signal, $\widehat{\mathbf{h}}_{\alpha}$, then the linear combination becomes an exact representation. In other words, the projection of the second signal onto the subspace $\mathcal{S}$ vanishes completely. It follows that, for lensed pairs, $\chi^2_{\rm{lens}}$ acquires the central $\chi^2$ distribution with $\langle \chi^2_{\rm{lens}} \rangle = 2$ and $\sigma_{\chisqrlns}^2 = 4$.

$\langle \chi^2_{\rm{lens}} \rangle$ for an unlensed pair grows as the square of the amplitude of the second signal, $|A_2|^2$. Moreover, it depends on the consistency (or lack thereof) between the two signals in the unlensed pair, quantified by $\| \Delta \mathbf{h} \|$. Chance astrophysical coincidences, which result in approximately similar source parameters for signals in the unlensed pair, will require larger SNRs to be confidently identified as unlensed. In other words, false alarms are reduced with increased signal amplitudes.

To get a better understanding of the ability of $\chi^2_{\rm lens}$ to discriminate between lensed and unlensed pairs, it is instructive to consider the following illustrative example. Let us suppose that we wish to correctly identify a pair of unlensed GW events at $c\%$ confidence. 

We note that the relevant distribution is the central $\chi^2$ with 2 degrees of freedom. This distribution is just the exponential distribution. We need the cumulative distribution function (CDF) of this distribution, which is
\be
P(z \leq Z) = 1 - e^{- Z/2} \,.
\ee
Setting $Z = \chi^2_{\rm crit}$ and $P (Z) = c/100$, also shown in Figure \ref{fig:chisq_cut}(left), we obtain,
\begin{equation}
    \chi^{2}_{\rm crit} = 2 \rm ln \left[ \frac{100}{100 - c} \right]\,.
\end{equation}

Then, on average, dismissal of an unlensed pair at $ > c\% $ confidence, as shown in the right panel of Figure \ref{fig:chisq_cut}, would translate to the following condition on the amplitude of the second signal

\begin{equation}
    \langle \chi^2_{\rm lens} \rangle_{\rm UL} > \chi^2_{\rm crit} \Rightarrow |A_2|^2 >  \frac{\chi^2_{\rm crit} - 2}{\| \Delta \mathbf{h} \|}\,.
\end{equation}

\subsection{Effect of mismatch}\label{Subsubsec:effect_of_mismatch}

In this section, we consider a more realistic scenario in which the second signal, $\widehat{\mathbf{h}}^{\rm s_2}$, and the corresponding trigger template, $\widehat{\mathbf{h}}_{0}^{\rm t_2}$, have different intrinsic parameters. If the SNR of the signal is reasonably large - say $\sim 15$ - the signal will generally lie in the immediate neighborhood of the trigger template. However, here we will assume that the signal lies in the immediate neighborhood, defined by the mismatch parameter $\epsilon \coloneq (1-\mathbb{M})$, of the trigger template. Consider a template $\widehat{\mathbf{h}}_{\phi}^{\rm t_2} = \widehat{\mathbf{h}}_{0}^{\rm t_2}e^{i\phi}$ with $\phi = \phi_{\rm max}$ such that the inner product $\left( \widehat{\mathbf{h}}^{\rm s_2}, \widehat{\mathbf{h}}_{\phi}^{\rm t_2}\right)_{\phi = \phi_{\rm max}}$ is maximized. Since there is a mismatch between the signal and the template, this inner product will be less than one.

Now, the signal can be written as,
\begin{equation}
    \widehat{\mathbf{h}}^{\rm s_2} = \widehat{\mathbf{h}}_{\phi_{\rm max}}^{\rm t_2} + \delta \mathbf{h}\,.
\end{equation}
From the previous section,
\begin{align}
   \left | \left(\widehat{\mathbf{h}}_{\phi_{\rm max}}^{\rm t_2}, \widehat{\Delta \mathbf{h}} \right) \right |
    &= \left |e^{-i\phi_{\rm max}} A_2\left (\widehat{\mathbf{h}}^{\rm t_2}_{0}, \widehat{\Delta \mathbf{h}} \right) \right | \nonumber \\
    &= A_2 \| \Delta \mathbf{h} \|.
\end{align}
Also, it has been shown in Ref. \cite{dhurandhar2017} that in the case of a mismatch between the signal and the template, the norm of $\delta \mathbf{h}$ is bounded above by
\begin{equation}
    \| \delta \mathbf{h} \| \leq \sqrt{2\epsilon}\,.
\end{equation}

Using the Cauchy-Schwartz inequality, it can be shown that the quantity $\left \lvert \left( \delta \mathbf{h}, \widehat{\Delta \mathbf{h}} \right)\right \rvert$ has an upper bound given by 

\begin{equation}
    \left| \left( \delta \mathbf{h}, \widehat{\Delta \mathbf{h}} \right)  \right| \leq 
    \left\| \mathbf{\delta h} \right\|.\left\| \widehat{\Delta \mathbf{h}} \right\| \leq \sqrt{2\epsilon}\,.
\end{equation}

\begin{figure}[ht!]
    \hspace*{-1.0cm}
    \centering
    \includegraphics[width=0.90\linewidth]{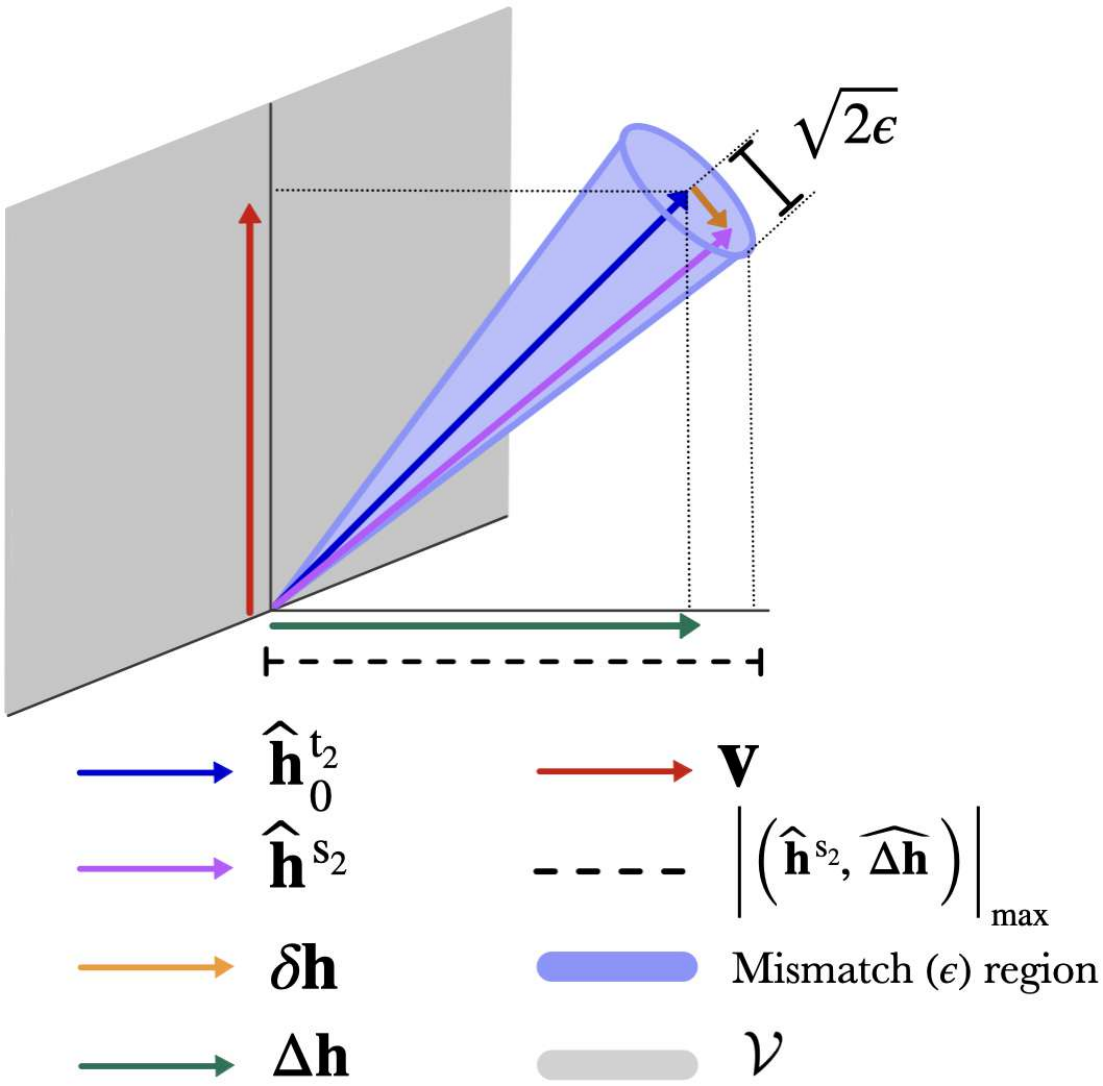}
    \caption{A visual representation of the bound presented in Eq.~\eqref{eq:bound_with_mismatch}. The approximate neighborhood space $\mathcal{V}$ has been represented by a plane (grey). The zero phase trigger template $\widehat{\mathbf{h}}_{0}^{\rm t_2}$ has a component $\mathbf{v}$ (red) in the space $\mathcal{V}$, while the other component $\Delta \mathbf{h}$ (dark green) being orthogonal to it. We assume that the signal $\widehat{\mathbf{h}}^{\rm s_2}$ (purple) lies in the immediate neighborhood (violet cone) of the trigger template. The length of the difference vector $\delta \mathbf{h}$ (orange) is bounded above by $\sqrt{2\epsilon}$. It can also be seen that the projection of the signal onto the orthogonal space has an upper bound (dashed line). }
    \label{fig:bound_illustration}
\end{figure}

Eventually, using the triangle inequality for complex numbers, it can be shown that the absolute value of the mean of correlation parameter $\Delta C$ is bounded above as,

\begin{align}\label{eq:bound_with_mismatch}
    \lvert \langle \Delta C \rangle \rvert &= \left| \left( A_2 \widehat{\mathbf{h}}^{\rm s_2}, \widehat{\Delta \mathbf{h}} \right) \right| \nonumber \\
    &= A_2\left| \left(\widehat{\mathbf{h}}_{\phi_{\rm max}}^{\rm t_2}, \widehat{\Delta \mathbf{h}} \right) + \left( \delta \mathbf{h}, \widehat{\Delta \mathbf{h}} \right) \right| \nonumber \\
    &\leq A_2 \left( \left| \left(\widehat{\mathbf{h}}_{\phi_{\rm max}}^{\rm t_2}, \widehat{\Delta \mathbf{h}}\right) \right| + \left| \left( \delta \mathbf{h}, \widehat{\Delta \mathbf{h}}\right) \right| \right) \nonumber \\
    &\leq A_2 \left( \| \Delta \mathbf{h} \|  + \sqrt{2\epsilon} \right)\,.
\end{align}

This bound has been illustrated in Figure~\ref{fig:bound_illustration}. The mean of the $\chi^2_{\rm{lens}}$ is then

\begin{equation}\label{eq:chisq_bound_with_mismatch}
    \langle \chi^2_{\rm{lens}} \rangle \leq  (A_{2})^{2} \left( \| \Delta \mathbf{h} \| + \sqrt{2\epsilon} \right)^2 + 2\,.
\end{equation} 

\begin{figure*}[p]
    \centering
    \includegraphics[scale=0.31]{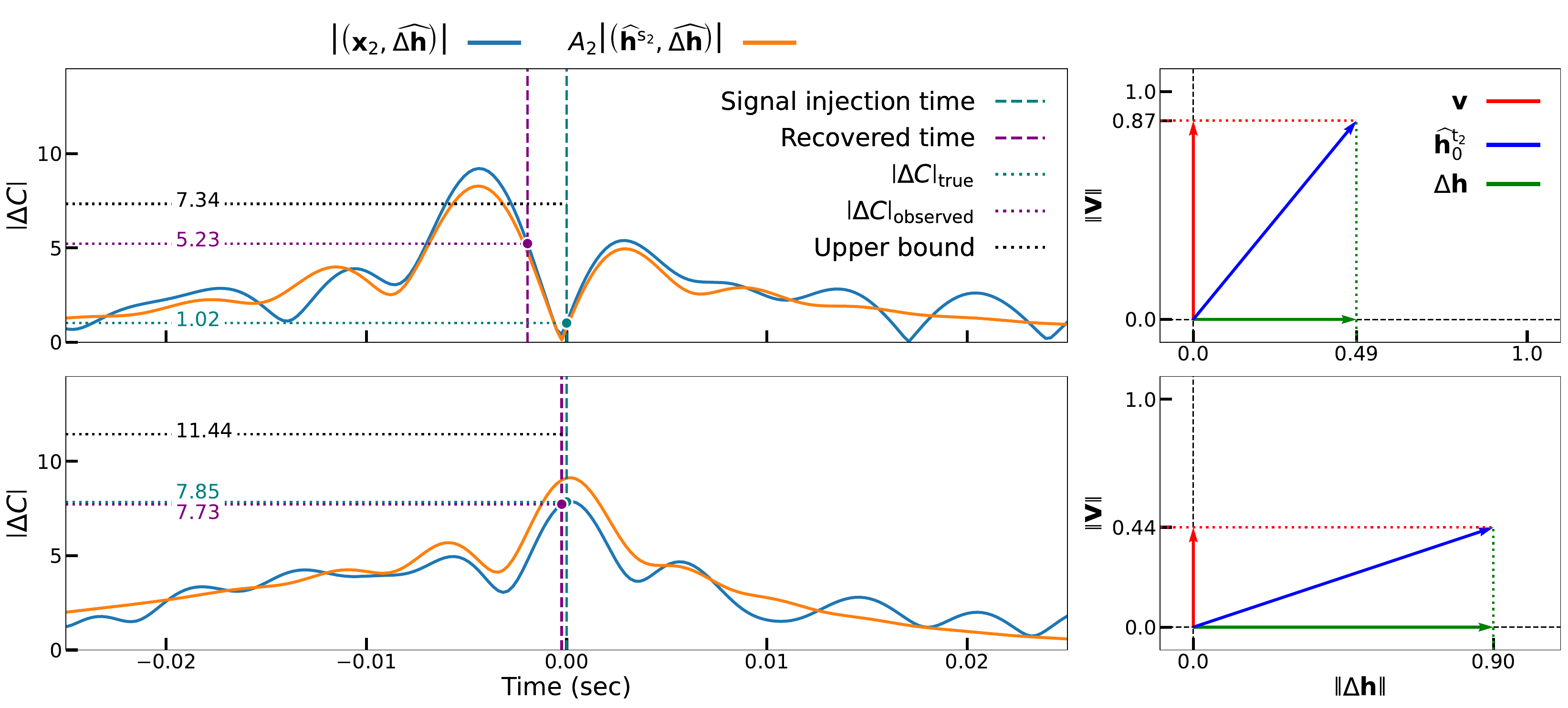}
    \caption{ The $|\Delta C(t) |$ time-series for the lensed (top left) and unlensed case (bottom left) with/without noise when $\Delta \mathbf{h} \perp \widehat{\mathbf{h}}^{\rm t_1}$. We observe a sharp dip at the merger time $t=0$ sec for the lensed case and a peak for the unlensed case. Due to an error in locating the signal merger time from matched filtering, we end up observing $|\Delta C|_{\rm observed}$(purple) instead of $|\Delta C|_{\rm true}$(teal). It should also be noted that $|\Delta C|_{\rm true}$ in both lensed/unlensed cases is well below the upper bound on $| \langle \Delta C \rangle|$ (black) derived in Eq.~\eqref{eq:bound_with_mismatch} for $\epsilon=0.97$. The top/bottom right plots show the parallel (red) and perpendicular (green) components of $\widehat{\mathbf{h}}_{0}^{\rm t_2}$ to the template $\widehat{\mathbf{h}}^{\rm t_1}$ for the lensed/unlensed cases, respectively. Both the trigger templates are more aligned towards each other in the lensed case compared to the unlensed case, quantified by $\| \Delta \mathbf{h}\|$.}
    \label{fig:corr_single}
\end{figure*}

\begin{figure*}[p]
    \centering
    \includegraphics[scale=0.31]
    {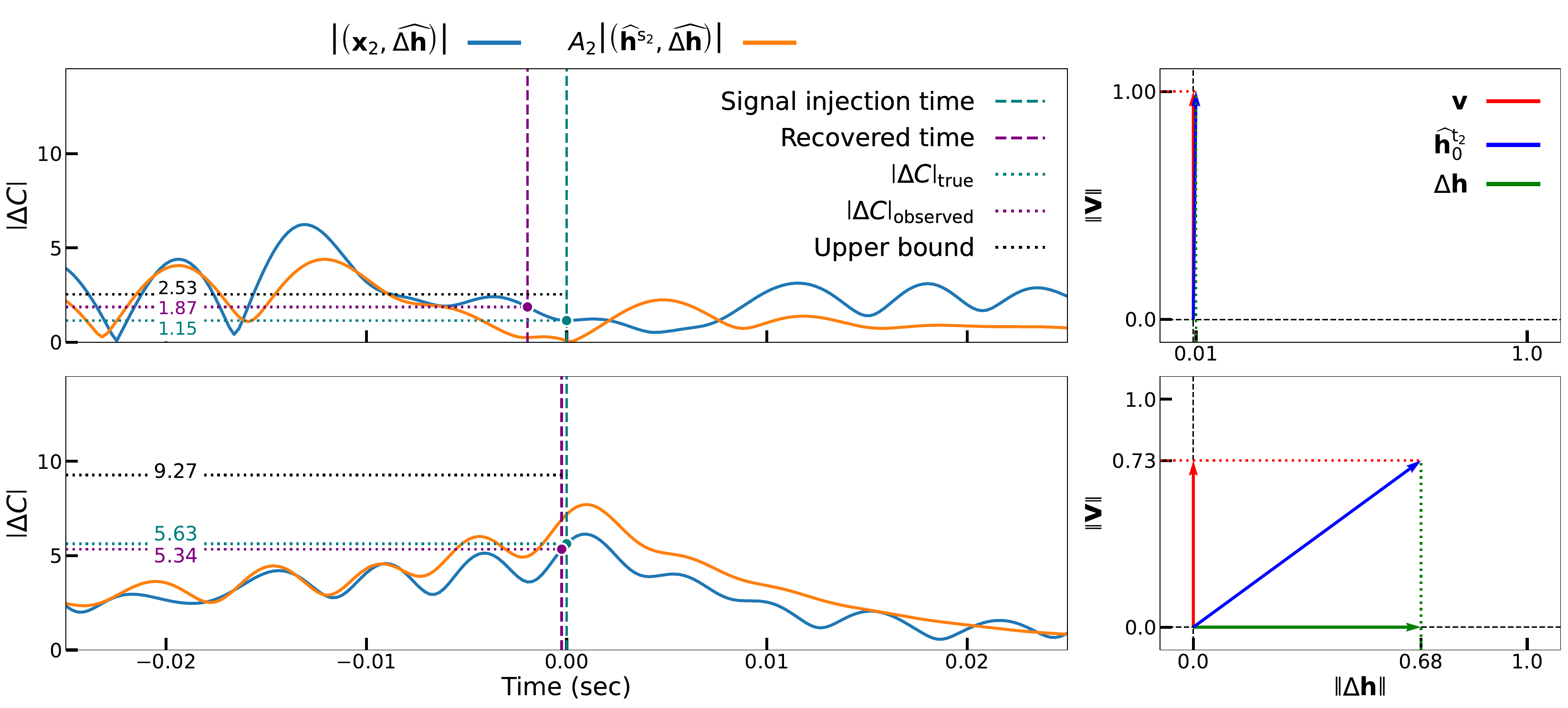}
    \caption{The $|\Delta C(t) |$ time-series for the lensed (top left) and unlensed case (bottom left) with/without noise when $\Delta \mathbf{h} \perp \mathcal{V}$. We observe a broader flattening of the correlation parameter time-series around the merger time $t=0$ sec for the lensed case and a slight depreciation of the peak for the unlensed case compared to the single template case in Figure \ref{fig:corr_single}. Again, the $|\Delta C|_{\rm true}$(teal) is smaller than the upper bound (black) on $| \langle \Delta C \rangle|$ for both lensed/unlensed cases. We also observe that $\widehat{\mathbf{h}}_{0}^{\rm t_2}$ is almost fully contained in the space $\mathcal{V}$ as $\| \mathbf{v} \| \approx 1$ for the lensed case (top right) making it a better discriminator compared to the single template case, although this also follows up with a relative increase in $\| \mathbf{v} \|$ for the unlensed case (bottom right) compared to the single template case in Figure \ref{fig:corr_single}.}
    \label{fig:corr_trunc}
\end{figure*}

We now want to highlight the importance of constructing a space orthogonal to the neighborhood space $\mathcal{V}$ rather than only the trigger template $\widehat{\mathbf{h}}^{\text{t}_1}_{0}$. Figures \ref{fig:corr_single} and \ref{fig:corr_trunc} show the $\left| \Delta C \right|$ time series for the lensed and unlensed cases with/without noise when: 
\begin{itemize}
    \item $\Delta \mathbf{h}$ is orthogonal to the single template $\widehat{\mathbf{h}}^{\rm t_1}_{0}$.
    \item $\Delta \mathbf{h}$ is orthogonal to the space $\mathcal{V}$.
\end{itemize}

\begin{figure*}[h!t]
    \centering
    \includegraphics[width=\textwidth]{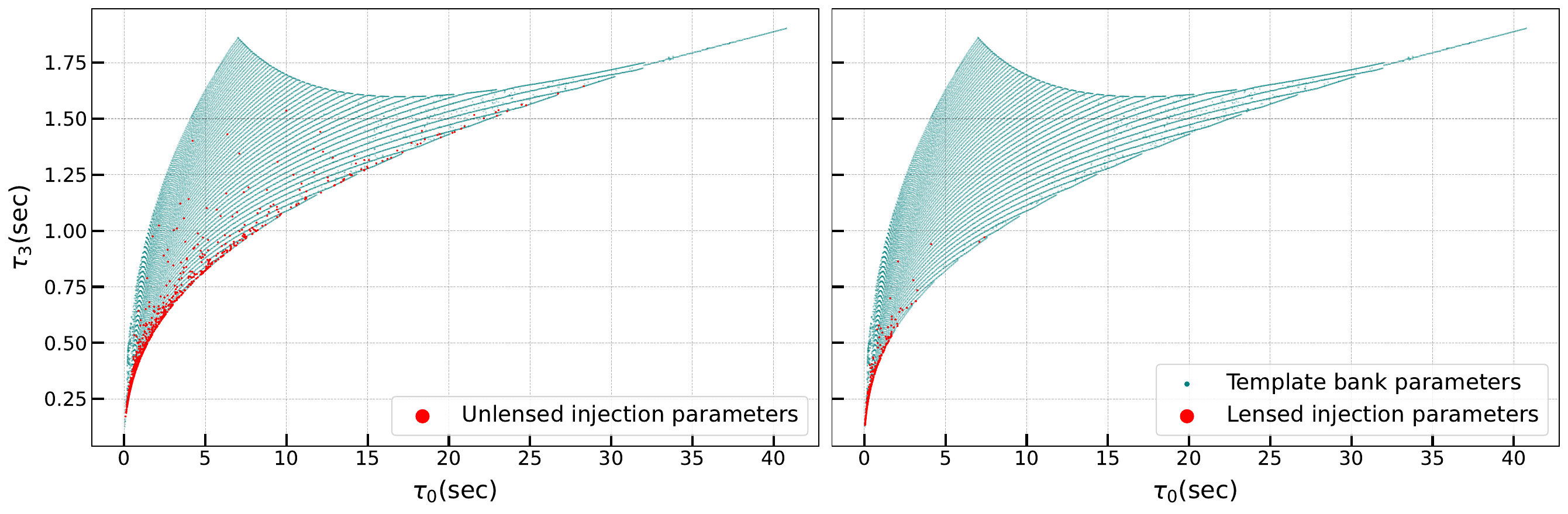}
    \caption{Chirp-time coordinates ($\tau_0$ and $\tau_3$) of unlensed injections (left) and lensed injections (right) taken from the DST dataset used in Ref.~\cite{Goyal:2021hxv}. The same coordinates pertaining to the templates in the bank are also plotted.}
    \label{fig:dst_inj_param}
\end{figure*}

Here, we show the case of a lensed event pair with signals having chirp mass ($M_{\rm c}$) of 36.9 $M_{\odot}$ and different amplitudes with $A_1 = 15$ and $A_2 = 10$, injected in different noise realizations. Similarly, for the unlensed case, we inject two signals with $M_{\rm c}^{(1)} = 36.9 \, M_{\odot}$ and $M_{\rm c}^{(2)} = 31.7 \, M_{\odot}$, having amplitudes $A_1 = 15$ and $A_2 = 10$, respectively, in different noise realizations. The signals have a merger time at $t = 0$ sec. 

Matched filtering of both the strain data with the templates in the template bank gives us the trigger templates with the highest trigger SNR. Figure \ref{fig:corr_single} (top and bottom left for lensed and unlensed cases, respectively) shows the case when $| \Delta C |$ time-series is evaluated with/without noise for the vector $\Delta \mathbf{h}$ defined as  

\begin{equation}
    \Delta \mathbf{h} = \widehat{\mathbf{h}}_{0}^{\rm t_2} - \left( \widehat{\mathbf{h}}^{\rm t_1}\,, \widehat{\mathbf{h}}_{0}^{\rm t_2}\right) \widehat{\mathbf{h}}^{\rm t_1}\,.
\end{equation}

In the single template case, the parallel component of $\widehat{\mathbf{h}}_{0}^{\rm t_2}$ turns out to be $\mathbf{v} = \left( \widehat{\mathbf{h}}^{\rm t_1}, \widehat{\mathbf{h}}_{0}^{\rm t_2}\right) \widehat{\mathbf{h}}^{\rm t_1}$. For the lensed case, we expect a dip at $t = 0$ sec since $\Delta \mathbf{h} \perp \widehat{\mathbf{h}}^{\rm t_1}$. We indeed observe a sharp dip at $t = 0$ sec, but due to uncertainty in locating the signal merger time from matched filtering, we may end up observing $\left| \Delta C \right|$ at $t \neq 0$ instead of the true one at $t = 0$. Similarly, we also show the unlensed case, where we expect a peak at the merger location. On the right-hand side of the figure, we show the parallel ($\mathbf{v}$) and perpendicular ($\Delta \mathbf{h}$) components of $\widehat{\mathbf{h}}_{0}^{\rm t_2}$ (top and bottom for lensed and unlensed case, respectively). An efficient discriminator would have $\| \mathbf{v} \|$  close to unity for the lensed case while being as small as possible for the unlensed case. 

Figure \ref{fig:corr_trunc} (top left) is for the case when $\Delta \mathbf{h} \perp \mathcal{V}$, where we find around 15 templates within $0.97$ match of $\widehat{\mathbf{h}}^{\rm t_1}$, which after SVD truncation is reduced to 5. We note that the dip around the merger time flattens out over a wider range around the signal merger time for the lensed case, hence mitigating the effect of getting a higher correlation for the lensed case due to errors in locating the merger time. Also note that $\| \mathbf{v} \| \approx 1$ implies that $\widehat{\mathbf{h}}_{0}^{\rm t_2}$ projects very well onto the space $\mathcal{V}$. Choosing a wider neighborhood with lower match criteria tends to increase the number of false alarms. For our simulations discussed in the next section, we perform our analysis with the neighborhood of $\mu = 0.97$, which turns out to be the optimal match criterion for the injection set used in this analysis.

\section{Results}\label{Sec:Results}

We test the performance of $\chi^{2}_{\rm lens}$ on the DST (DataSet Training) dataset (see Appendix \ref{append_table} for injection parameters) used in Ref.~\cite{Goyal:2021hxv}. This dataset consists of $\sim 300$ lensed injection pairs and $1000$ unrelated injections. By combining the unrelated injections pairwise, we construct roughly half a million unlensed pairs for our analysis, resulting in a lensed-to-unlensed ratio of about $1{:}1600$. The distribution of injection parameters, along with the template bank parameters, is shown in Figure~\ref{fig:dst_inj_param}. The prescription to generate the injection dataset has been detailed in Appendix A of \cite{Haris:2018vmn}. The template bank used in our analysis is a 3.5 post-Newtonian (PN) geometric bank~\cite{babak2006template}. Since we consider nonspinning injections in our analysis, we visualize the intrinsic parameters in \textit{chirp time} $(\tau)$ coordinates~\cite{Cokelaer_2007} defined as follows:

\begin{subequations}
    \begin{align}
        \tau_{0} &= \frac{5}{256\pi\eta f_{\rm low}}\left( \pi M_{\rm tot} f_{\rm low} \right)^{-5/3}\,,\\
        \tau_{3} &= \frac{1}{8\eta f_{\rm low}} \left( \pi M_{\rm tot} f_{\rm low} \right)^{-2/3}\,.
    \end{align}
\end{subequations}

\begin{figure*}[tb]
    \centering
    \includegraphics[width=\textwidth]{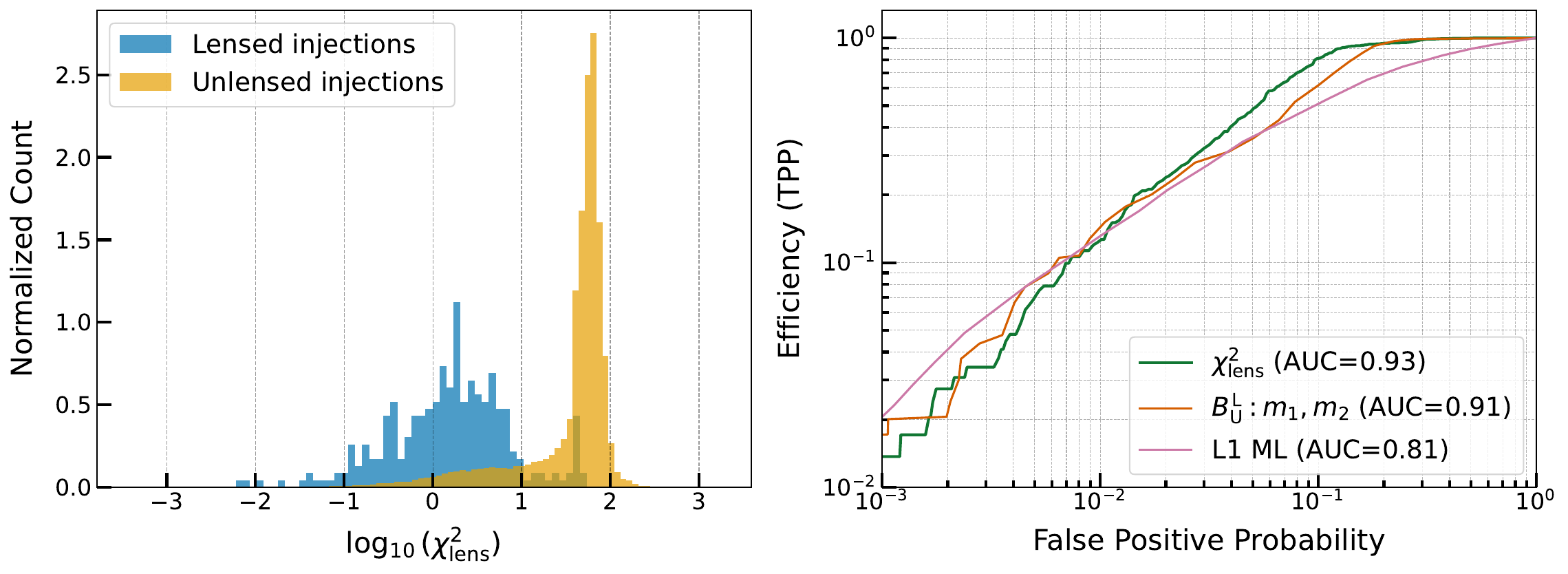}
    \caption{(Left) The histograms of $\log_{10} \left( \chi^{2}_{\rm lens} \right)$ computed for lensed and unlensed signals in the DST dataset. (Right) ROC curves for $\chi^{2}_{\rm lens}$ statistic tested on DST dataset (green), the $m_1, m_2$ $B_{\rm U}^{\rm L}$ statistic (orange) from Ref.~\cite{Haris:2018vmn} and single detector machine learning classifier (pink) from Ref. \cite{Goyal:2021hxv}. The figure shows that the performance of $\chi^{2}_{\rm lens}$ (AUC = 0.93) is comparable to that of $m_1, m_2$ $B_{\rm U}^{\rm L}$ (AUC = 0.91), while it is significantly better than single detector machine learning (ML) classifier (AUC = 0.81).}
    \label{fig:dst_roc_hist}
\end{figure*}

Here $\tau_0$ is the Newtonian time of coalescence of the signal and $\tau_3$ corresponds to the 1.5 PN order. $\eta$ is the ratio of the reduced mass to the total mass $M_{\rm tot}$ calculated in the detector frame. For all computations, we use the zero-detuned high-power PSD of Advanced LIGO at design sensitivity, with lower frequency cutoff $f_{\rm low} = 15$ Hz, as implemented in \textsc{PyCBC} \cite{pycbc}. All waveforms for both the signal and the templates are generated using the \textsc{IMRPhenomD} waveform approximant \cite{husa2016frequency, khan2016frequency}. Since $\chi^{2}_{\rm lens}$ is a single detector test, we \textit{rescale the luminosity distances} for all injections so that the signal amplitudes $A_{1}$ and $A_2$ are at least 8. 

We assess the performance of the $\chi^{2}_{\rm lens}$ discriminator by comparing its receiver operator characteristic (ROC) curve with that of the posterior overlap statistic $B_{\rm U}^{\rm L}$ \cite{Haris:2018vmn}, which uses the posteriors of the component masses ($m_1, m_2$). A ROC curve for the single detector machine learning classifier \cite{Goyal:2021hxv} is also shown. The ROC curve comparison, along with a histogram of $\chi^{2}_{\rm lens}$ statistic evaluated for each of the lensed/unlensed event pairs, is shown in Figure \ref{fig:dst_roc_hist}. We find that the performance of $\chi^{2}_{\rm lens}$ is comparable to the $B_{\rm U}^{\rm L}$ statistic, quantified by Area Under the Curve (AUC) of the ROC curve, which for $\chi^{2}_{\rm lens}$(AUC=0.93) is marginally larger than $B_{\rm U}^{\rm L}$(AUC=0.91) and non-trivially larger than the single detector machine learning classifier (AUC=0.81).

\begin{figure*}[tb]
    \centering
    \includegraphics[width=\textwidth]{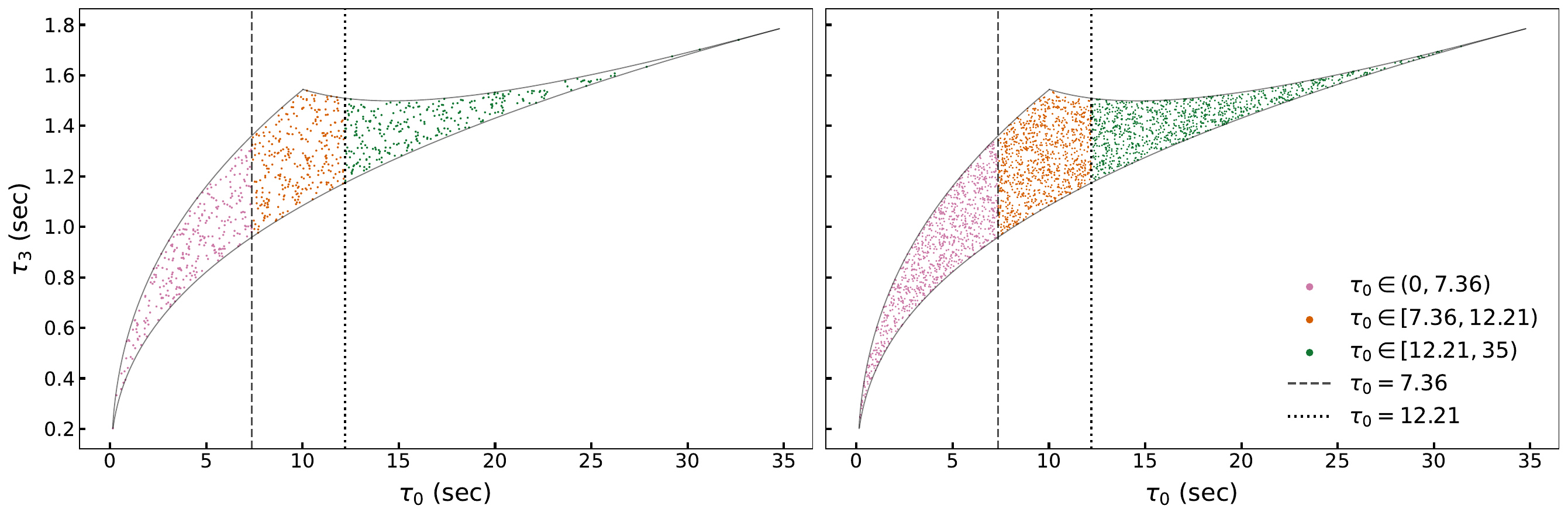}
    \caption{Chirp-time coordinates of lensed and unlensed injections for testing the performance of $\chi^{2}_{\rm lens}$ across three different $\tau_0$ bins. Each $\tau_0$ bin contains 300 points for lensed (left) and 1000 points for the unlensed (right) injections sampled from a uniform distribution.}
    \label{fig:tau0_binwise}
\end{figure*}

\begin{figure*}[h!t]
    \centering
    \includegraphics[width=\textwidth]{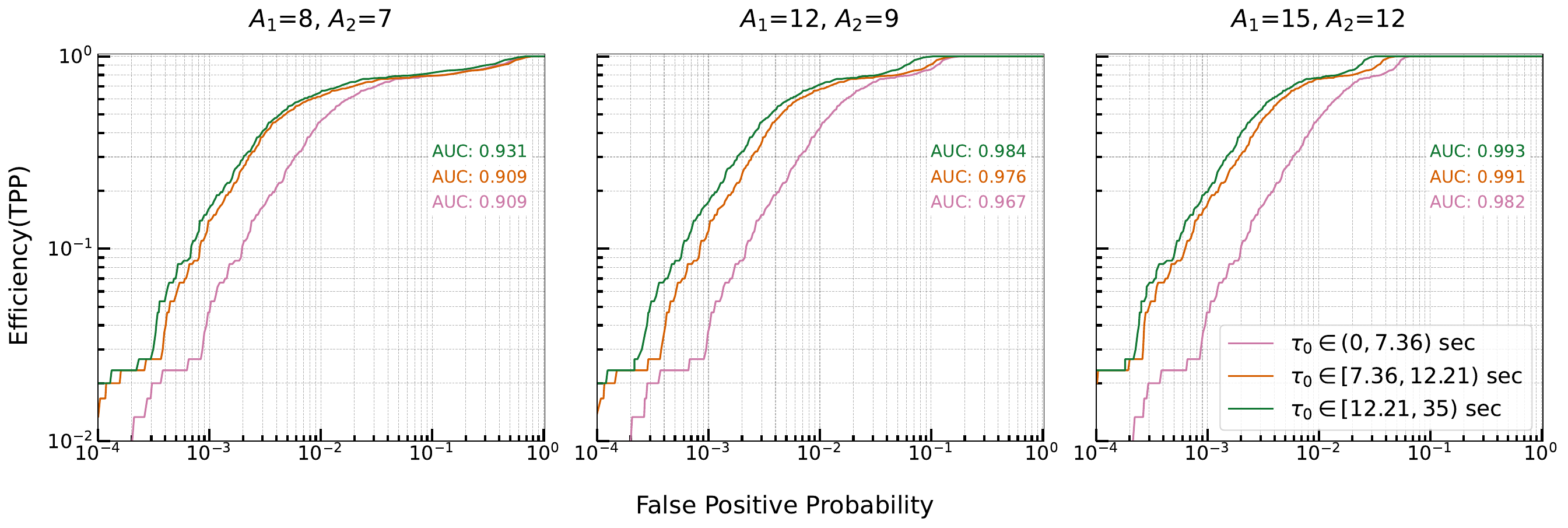}
    \caption{ROC curves for the $\chi^{2}_{\rm lens}$ statistic tested for performance across different rescaled optimal SNR combinations for the signals when $\left( A_1, A_2 \right)$ are (8, 7) (left), (12, 9) (middle) and (15, 12) (right), which are further tested across different $\tau_0$ bins. As a measure of the performance of $\chi^{2}_{\rm lens}$ for different SNR combinations and $\tau_0$ bins, we report AUC for each ROC curve. We find a consistent pattern that increasing the SNR of the pair for a given $\tau_0$ bin tends to improve the ROC curve, indicated by an increase in AUC. Also, for a given SNR combination of the pair, signals in high $\tau_0$ bins tend to perform better due to an increase in the duration of the signals.}
    \label{fig:roc_binwise}
\end{figure*}

We further test the performance of $\chi^{2}_{\rm lens}$ statistic across various $\tau_0$ bins for varying SNR combinations. We define a parameter space region with $m_1, m_2 \in \left[ 5.5 M_{\odot}, 140 M_{\odot}\right]$ and the mass ratio $q \leq 5$ \footnote{$q \equiv m_1/m_2$, $m_1 \geq m_2$.}, which in $\tau$ coordinates has been shown in Figure \ref{fig:tau0_binwise}. We divide the whole region into three equal areas and draw 300 and 1000 uniform samples of $\tau_0$ - $\tau_3$ parameters for the lensed and unlensed analysis, respectively, from each $\tau_0$ bin. In each bin, we test the performance of $\chi^{2}_{\rm lens}$ by rescaling the luminosity distance so that the amplitudes ($A_{1}, A_{2}$) for signals in the event pair are (8, 7), (12, 9) and \break (15, 12). Figure \ref{fig:roc_binwise} shows the ROC curves for each SNR combination across different $\tau_0$ bins. We observe a consistent trend where the performance of the $\chi^{2}_{\rm lens}$ classifier, quantified by the AUC, improves as the duration of the signal ($\tau_0$) increases. Also, the performance improves with an increase in the SNRs of both signals.

\section{Summary and Outlook}\label{Sec:Conclusions}

A number of methods to search for signatures of gravitational lensing have been developed and deployed on LVK data (see, e.g., Ref.~\cite{LIGOScientific:2023bwz}, and references therein). To date, no confirmed detection of GW lensing has been reported. Nevertheless, the expectation is that $0.1\% - 1\%$ of the GWs detectable by the LVK detector network will be lensed \cite{wierda2021beyond, Xu:2021bfn}. Thus, the first detection of strongly lensed GWs by galaxy or cluster scale lenses is anticipated to happen before the end of O5. 

Our work proposes a novel $\chi^2$ statistic to identify strongly lensed GW pairs and dismiss unlensed pairs. The method exploits the expected phase evolution consistency between the events in the lensed pair, which, for unlensed pairs, should be absent. In particular, it tailors the unified $\chi^2$ statistic delineated in \cite{dhurandhar2017} to the problem at hand. Using the template of the louder signal and its immediate neighborhood in the template bank, a vector orthogonal to the space spanned by the said template and its neighborhood is constructed. The data pertaining to the second signal is then projected onto the orthogonal vector to construct the $\chi^2_{\rm lens}$ statistic. 

We assess the performance of $\chi^2_{\rm lens}$ on an in-part astrophysically motivated dataset consisting of lensed and unlensed pairs. We find that our method provides detection efficiencies, at low false-positive probabilities, that are comparable to other standard lensing searches involving Bayesian and machine learning methods. We also find that $\chi^2_{\rm lens}$ systematically improves its ability to separate lensed pairs from unlensed pairs with increasing SNR and in-band signal duration. We ascertain that, indeed, the use of the neighborhood around the template of the louder signal, in addition to the template itself, significantly improves the performance of $\chi^2_{\rm lens}$ with respect to when only the template itself is used. This fact was conjectured, in the context of unified $\chi^2$, in \cite{dhurandhar2017}, and has now been exemplified in this work. 

Our method has two advantages over standard lensing searches. The first is speed. This statistic can be evaluated in a fraction of a second per candidate pair and is readily implementable in a matched filter search pipeline. This not only allows for low-latency dissemination of GW lensing information but also significantly alleviates the computational burden required by large-scale Bayesian methods \cite{Janquart:2022wxc, Jana:2024dhc, Lo2023}. Moreover, $\chi^2_{\rm lens}$ provides a suitable alternative for the rapid (preliminary) identification of subthreshold image counterparts to superthreshold GW events \cite{Goyal:2023lqf}. Interesting candidate pairs unearthed with $\chi^2_{\rm lens}$ can then be further followed up with large-scale Bayesian methods. In fact, we expect that $\chi^2_{\rm lens}$ can be a valuable and easily integrable addition to subthreshold search pipelines such as TESLA \cite{Li:2019osa, Li:2023zdl}. 

The second is the interpretation. Arguably, unlike most of the lensing search methods currently available, in stationary Gaussian noise, the statistics of $\chi^2_{\rm lens}$ are fully understood. In stationary Gaussian noise, $\chi^2_{\rm lens}$ for lensed pairs follow a central $\chi^2$ distribution, while unlensed pairs follow a non-central $\chi^2$ distribution whose mean is proportional to the squared SNR of the second (weaker) signal. This, perhaps, is the most attractive feature of $\chi^2_{\rm lens}$. However, we make an underlying assumption that the signal model used for searches captures all the physics of the true signal. In case there are other physical effects present in the signal which are not accounted for in the templates while performing the searches, it will limit the performance of $\chi^{2}_{\mathrm{lens}}$.

It is worth reiterating that our method essentially probes phase evolution consistency between pairs of GW events. In general, lensing in the geometric optics regime is expected to exhibit phase evolution consistency between the images. However, there are exceptions where the images violate this consistency. For example, it has been shown that Type II images of GW sources with significant higher-mode content will have a phase morphology that is distinct from that of the GWs produced by the source \cite{wang2021identifying, janquart2021identification, Vijaykumar:2022dlp, Taylor:2024yjt}. On the other hand, Type I and III images will have morphologies identical to that of the source. Consequently, phase-evolution consistency between pairs of images, where exactly one of them is a Type II image of a GW source with significant higher-mode content, will be violated. Thus, all methods that look for phase-evolution consistency between GW event pairs, including ours, the traditional posterior overlap statistic method \cite{Haris:2018vmn, Goyal:2023lqf, Barsode:2024zwv}, machine learning methods \cite{Goyal:2021hxv, Magare:2024wje}, and cross-correlation methods \cite{Chakraborty:2024net}, will not work for said exceptions, without adequate modifications. We have also made an assumption that the strongly lensed signals do not undergo any extra amplitude and phase modulation due to the wave optics effect caused by micro-lenses embedded within the macro-lens, which may result in phase evolution inconsistency \cite{diego2019observational, seo2025residual, mishra2024exploring, shan2024microlensing, yeung2023detectability} between the strongly lensed images. 

We point out that, besides phase-evolution consistency, additional features, such as sky localization, can help dismiss unlensed pairs. The expectation is that strongly lensed pairs of GW images should have arc or sub-arsecond angular separations, assuming galaxy or cluster-scale lenses. Given that localisation sky areas for current GW detector networks can span ten to hundreds of square degrees, lensed pairs should have skymaps with large overlaps. On the other hand, there is no such expectation for pairs of events that are unrelated. Thus, a quantitative measure of this overlap, using Eq.~\ref{eq:blu} \cite{Haris:2018vmn, Goyal:2021hxv}, could be used, in tandem with $\chi^2_{\rm lens}$, to construct a joint-discriminator. It might also be possible to incorporate prior information on expected image time delays and relative image magnifications into the joint-discriminator, along the lines pursued in \cite{more2022improved}. We leave this for future work.  

\section*{Acknowledgements}
We thank Bhooshan Gadre for their reading of the manuscript. We also thank Shreejit Jadhav and Anirban Kopty for useful discussions. SG's research was supported by the University Grants Commission, Government of India. SJK gratefully acknowledges support from the Science and Engineering Research Board (SERB) Grant SRG/2023/000419. KS acknowledges support from the National Science Foundation grant (PHY-2309240). All calculations and simulations were performed on the Sarathi computing cluster at IUCAA. 

\textit{Software}: \texttt{PyCBC} \citep{pycbc}, \texttt{NumPy} \citep{vanderWalt:2011bqk}, \texttt{SciPy} \citep{Virtanen:2019joe}, \texttt{astropy} \citep{2013A&A...558A..33A, 2018AJ....156..123A}, \texttt{Matplotlib} \citep{Hunter:2007}, \texttt{jupyter} \citep{jupyter}.

\appendix

\section{The Unified $\chi^2$ formalism}
\label{append_unified_chisq}

The $\chi^2$ discriminator maps a vector in $\D$ to a positive real number. It is defined so that its value on the signal is zero and, in Gaussian noise, has a $\chi^2$ distribution with a reasonable number of degrees of freedom, $p$. 
If a template $\h$ is triggered, then the $\chi^2$ for $\h$ is defined by choosing a finite-dimensional subspace $\myS$ of dimension $p$ that is orthogonal to $\h$, i.e., for any arbitrary vector $\y \in \myS$, we must have $(\y| \h) = 0$. Then $\chi^2$ for the template $\h$ is defined as just the square of the $L_2$ norm of the data vector $\x$ projected onto $\myS$. Specifically, we perform the following operations. Take a data vector $\x \in \D$ and decompose it as
% \be
% \x = \x_{\myS} + \x_{\myS^\perp} \,,
% \ee
\begin{equation}
    \mathbf{x} = \mathbf{x}_{\mathcal{S}} + \mathbf{x}_{\mathcal{S}^\perp} \,,
\end{equation}
where $\myS^{\perp}$ is the orthogonal complement of $\myS$ in $\D$. $\x_{\myS}$ and $\x_{\myS^\perp}$ are projections of $\x$ into the subspaces $\myS$ and $\myS^{\perp}$, respectively. We may write $\D$ as a direct sum of $\myS$ and $\myS^{\perp}$, that is, $\D = \myS \oplus \myS^{\perp}$. Then the required statistic $\chi^2$ is,
% \be
% \chi^{2} (\x) = \| \x_{\myS} \|^2 \,.
% \ee
\begin{equation}
    \chi^{2} (\mathbf{x}) = \| \mathbf{x}_{\mathcal{S}} \|^2 \,.
\end{equation}
The $\chi^2$ statistic so defined has the following properties. Given any orthonormal basis of $\myS$, say $\e_{\a}$, with $\a = 1, 2, ..., p$ and $(\e_{\a}| \e_{\b}) = \delta_{\a \b}$, we obtain the following:

\begin{enumerate}
    \item For a general data vector $\x \in \D$, we have
    % \be
    % \chi^2 (\x) = \| \x_{\myS} \|^2 = \sum_{\a = 1}^p  \left\lvert (\x | \e_{\a}) \right\rvert^2 \,. 
    % \ee   
    \begin{equation}
        \chi^2 (\mathbf{x}) = \| \mathbf{x}_{\mathcal{S}} \|^2
        = \sum_{\alpha = 1}^{p} \left| \left( \mathbf{x} \,\middle|\, \mathbf{e}_{\alpha} \right) \right|^2 \,.
    \end{equation}
    \item Clearly, $\chi^2 (\h) = 0$ because the projection of $\h$ onto the subspace $\myS$ is zero, i.e., $\h_{\myS} = 0$. 
    
    \item Now, 
    the noise $\n$ 
    is taken to be stationary and Gaussian, with PSD $S_{\rm n}(f)$ and mean zero. Therefore, the following relation is valid:
    % \be
    % \chi^2 (\n) = \| \n_{\myS} \|^2 = \sum_{\a = 1}^p |(\n| \e_{\a})|^2 \,.
    % \ee
    \begin{equation}
        \chi^2 (\mathbf{n}) = \| \mathbf{n}_{\mathcal{S}} \|^2
        = \sum_{\alpha = 1}^{p} \left| \left( \mathbf{n} \,\middle|\, \mathbf{e}_{\alpha} \right) \right|^2 \,.
    \end{equation}

    The random variables $(\n| \e_{\a})$ are independent and Gaussian distributed, with zero mean and unit variance. This is because $\langle (\e_{\a}| \n) (\n| \e_{\b}) \rangle = (\e_{\a}| \e_{\b}) = \delta_{\a \b}$, where the angular brackets denote ensemble average (see Ref.~\citep{Creighton:2011zz} for proof). Thus, $\chi^2 (\n) $ possesses a $\chi^2$ distribution with $p$ degrees of freedom. 
\end{enumerate}

One is free to choose any orthonormal basis of $\myS$. In an orthonormal basis, the statistic is manifestly $\chi^2$ since it can be written as a sum of squares of independent Gaussian random variables, with mean zero and variance unity.
\par

In the context of CBC searches, however, we have a family of waveforms that depends on a set of parameters $\vec{\Theta}$. These include intrinsic parameters, $\vec{\theta}$, such as the masses and spins of the binary components, as well as kinematical parameters, such as the phase and time of coalescence. The templates corresponding to these waveforms are normalized, i.e., $\| \h (\vec{\Theta}) \| = 1$. Then the templates trace out a submanifold $\P$ of $\D$, called the {\it signal manifold}. We now associate a $p$-dimensional subspace $\myS$ orthogonal to the template $\h (\vec{\Theta})$ at each point of $\P$. Thus, we have a $p$-dimensional vector space ``attached'' to each point of $\P$. When done smoothly, this construction results in a vector bundle. We have, therefore, found a very general mathematical structure for the $\chi^2$ discriminator. It is easily shown that the traditional $\chi^2$ discriminator \cite{Allen2005} falls under the class of {\it unified} $\chi^2$ \cite{dhurandhar2017}.
\par

\section{Derivations}
\label{append_A}

The Hilbert space we are dealing with is $L^2[\fmn, \fmx]$ with the weighted measure $d \mu = df/S_{\mathrm{n}} (f)$, which we have already denoted by $\H$. Given time series data that is discretely sampled, we have, say, $M$ number of frequency bins in the frequency range $[\fmn, \fmx]$. Then, the Hilbert space is just $C^M$ with a weighted measure $\mu$ as defined. This is consistent with the practice of using a one-sided PSD with only positive frequencies. 
\par

Two different scalar products have been defined by Eqs.~\eqref{eq:scalar_gw} and~\eqref{eq:scalar_u}, namely, $(\x | \y)$ and $(\x, \y)$ for two vectors $\x$ and $\y$ on the vector space. We may easily verify that these two scalar products are related by
\bea
\label{eq:rel_sclr_prdct_1}
(\x | \y) &=& \hf [(\x, \y) + (\x, \y)^\star] \,, \\
(\x, \y) &=& (\x | \y) - i (\x, i\y) \,.
\label{eq:rel_sclr_prdct_2}
\eea
Thus, both scalar products are contentwise equivalent. So, from the mathematical point of view, we are dealing with two Hilbert spaces with their own scalar product, the vector space being the same. Eqs.~\eqref{eq:rel_sclr_prdct_1} and~\eqref{eq:rel_sclr_prdct_2} show that the two Hilbert spaces are isomorphic. Also, we notice from Eq.~\eqref{eq:rel_sclr_prdct_1} that,
\be
\| \x \|_|^2 = (\x | \x) = (\x, \x) = \|\x \|_,^2 \,,
\label{eq:norm}
\ee
that is, both scalar products yield the same norm for any vector $\x \in \H$.

\par

Secondly, we note that, since we are considering complex vectors, they contain full information about the negative frequency components of the signal and also the phase information. Thus, we write a signal with coalescence phase zero as $\th_0 (f) = u(f) + i v(f)$, where $f > 0$, and $u(f), v(f)$ real. Then the following relations hold:
\bea
\th_0(-f) &=& \th_0^*(f) = u(f) - i v(f) \,, \\
\th_{\pi/2} (f) &=& i \th_0 (f) = i u(f) - v(f)  \,, \\
\th_{\pi/2} (-f) &=& \th_{\pi/2}^* (f) = -i u(f) - v(f) \,.
\eea
The above relations show that given $\th_0(f)$ as a complex vector for $f > 0$, we can obtain the time domain signal $h_0 (t)$ as well as $h_{\pi/2} (t)$ pertaining to the two coalescence phases $0$ and $\pi/2$.\\

\section{Injection Parameters}
Table \ref{tab:summary-L} and \ref{tab:summary-UL} show the range of lensed and unlensed parameters, respectively, where $m_{1, 2\mathrm{z}}$ are the detector-frame masses and $t_{\mathrm{d}}$ (for lensed injections) is the time-delay between time of arrival of different images.
\label{append_table}

    \begin{table}[h!]
    \centering
    \caption{Lensed Signal Injection Parameters}
    \label{tab:summary-L}
    \begin{tabular}{l c c c}
    \hline
    \textbf{Parameter} & \textbf{Range} \\
    \hline
    $m_{1\mathrm{z}}$ & [15.8, 383.0]$M_{\odot}$ \\
    $m_{2\mathrm{z}}$ & [11.0, 258.0]$M_{\odot}$ \\
    $A_1$ & [5.2, 61.3] \\
    $A_2$ & [5.0, 60.0] \\
    $t_{\mathrm{d}}$ & [0.02, 23003911.1] sec\\
    $A_1/A_2$ & [1, 2.5] \\
    \hline
    \end{tabular}
    \end{table}
    \begin{table}[h!]
    \centering
    \caption{Un-lensed Signal Injection Parameters}
    \label{tab:summary-UL}
    \begin{tabular}{l c c c}
    \hline
    \textbf{Parameter} & \textbf{Range} \\
    \hline
    $m_{1\mathrm{z}}$ & [6.4, 240] $M_{\odot}$ \\
    $m_{2\mathrm{z}}$ & [5.5, 159] $M_{\odot}$ \\
    \hline
    \end{tabular}
    \end{table}

\bibliography{references}

\end{document}